# Differential Fault Analysis on A.E.S.

P. Dusart,* G. Letourneux,† O. Vivolo‡

01/10/2002

**Abstract**

We explain how a differential fault analysis (DFA) works on AES 128, 192 or 256 bits.

## Contents

## 1 Introduction

In September 1996, Boneh, Demillo, and Lipton [4] from Bellcore announced a new type of cryptanalytic attack which exploits computational errors to find cryptographic keys. Their attack is applicable to public key cryptosystems such as RSA, excluding secret key algorithms. In [3], E. Biham & A.Shamir extend this attack to various secret key cryptosystems such as DES, and call it Differential Fault Analysis (DFA). They applied the differential cryptanalysis to Data Encryption Standard (DES) in case of hardware fault model.

We further assume that the attacker is in physical possession of the tamperproof-device, so that he can repeat the experiment with the same cleartext and key but without applying the external physical effects. As a result,

---

*LACO (UMR CNRS n°6090), Faculté des Sciences, 123, avenue Albert THOMAS, 87060 Limoges, France
†E.D.S.I. Atalis 1, 1, rue de Paris, 35510 Cesson-Sévigné, France
‡development@edsi-smartcards.com.fr, E.D.S.I. Atalis 1, 1, rue de Paris, 35510 Cesson-Sévigné, France

he obtains two ciphertexts derived from the same (unknown) cleartext and key, where one of the ciphertexts is correct and the other is the result of a computation corrupted by a single error during the computation.

The DES used a 56-bits key which seems to be too short for future. Hence a mondial competition between secret key cryptosystems has been realized. The requirements of this standard is to replace DES standard: a symmetric cryptosystem with 128 to 256 key sizes, which can be easily implemented in hardware. On Oct. 2, 2000, NIST choose Rijndael to be the Advanced Encryption Standard (AES). AES uses a 128, 192 or 256 bits key with a 128 bits input message. It works on bytes with an algebraic structure which is the finite field $GF(2^8)$. Rijndael has been choosed by NIST for its resistance to linear and differential cryptanalysis [5].

The major critique of DFA was the practical feasibility of the theory. But some authors [2] have designed practical experimentations of this kind of attack with the possibility to inject the fault in a temporal windows which can be clearly related with program running process. By exposing a sealed tamperproof device such as a smartcard to certain physical effects (e.g., ionizing or microwave radiation), one can induce with reasonable probability a fault at a short random bit location in one of the registers at some intermediate stage in the cryptographic computation. In practice, the perturbation can change more than one bit. We assume that it can change up to one byte anywhere between the last two MixColumn operations of AES.

For DFA on DES, the attacker knows the differential input and output of the touched SBox. For AES, contrary to DES, we don't have the value of the differential fault $\varepsilon$ which could be obtained by considering the left part of the final DES state at round 16. For AES, if we consider a single fault before the SubBytes transformation, we can't go back to the key (There are 127 possibilities of the injected fault and 256 possibilities of a single byte of the round key, so the AES is protected against classical differential analysis.).

When the injected fault is becoming several induced faults (at least two) occuring in different bytes of the state, we can intersect each set of possible induced faults (the cardinal of intersection is lower than 63) and so we find a set of possible values (at most 128) for several bytes of the last subkey.

Further we find the last subkey with enough pairs of correct cipher/fault cipher. Once known this subkey is, we can find easily the key. For the sake of simplicity, we first assume that the first byte of the state before the MixColumn transformation of the nine round is replaced by a unknown value. The induced fault is going to be propagated by the MixColumn and spread over four bytes of the state. There is a linear relation between the four induced faults. For each byte is possible to find a set of possible value of induced fault, and then a set of possible values for the roundkey 10.

In this paper, we show that AES is sensitive to Fault Analysis. We have implemented this attack on a personal computer. Our analysis program found the full AES-128 key by analysing less than 50 ciphertexts.

## 2 The description of the AES

In this article, we use a description slightly different from the original AES submission FIPS PUB 197 [1]. We describe AES using matrix on $GF(2^8)$ but we try to keep the notations of [1].

The AES is a block cipher with block length to 128 bits, and support key lengths $N_k$ of 128, 192 or 256 bits. The AES is a key-iterated block cipher : it consists of the repeated application of a round transformation on the state. The number of rounds is denoted $N_r$ and depends on the key length ($Nr = 10$ for 128 bits, $Nr = 12$ for 192 bits and $N_r = 14$ for 256 bits).

The AES transforms a state, noted $S \in M_4(GF(2^8))$ , (i.e. $S$ is a matrix 4x4 with its coefficients in $GF(2^8)$) to another state in $M_4(GF(2^8))$. The key $K$ is expanded into $N_r + 1$ subkeys noted $K_i \in M_4(GF(2^8))$ ($i = 0, 1, \ldots, N_r$).

A round of an encryption with AES is composed of four main operations :

1. AddRoundKey
2. MixColumn
3. SubBytes
4. ShiftRows

### 2.1 Representation chosen for $GF(2^8)$

The representation chosen in [1] of $GF(2^8)$ is $GF(2)[X]/<m>$, where $<m>$ is the ideal generated by the irreducible polynomial $m \in GF(2)[X]$, $m = x^8 + x^4 + x^3 + x + 1$.

## 2.2 Notations used in this article

We use four notations for representing an element in $GF(2^8)$, which are equivalent to one another:

1. $x^7 + x^6 + x^4 + x^2$, the polynomial notation
2. $\{11010100\}_b$, the binary notation
3. 'D4', the hexadecimal notation
4. 212, the decimal notation

## 2.3 AddRoundKey for $i^{th}$ round

The AddRoundKey transformation consists of an addition of matrix in $M_4(GF(2^8))$ between the state and the subkey of the $i^{th}$ round. We denote by $S_{i,A}$ the state after the $i^{th}$ AddRoundKey.

$$\begin{array}{rcl} M_4(GF(2^8)) & \longrightarrow & M_4(GF(2^8)) \\ S & \longmapsto & S_{i,A} = S + K_i \end{array}$$

## 2.4 SubByte for $i^{th}$ round

The SubByte transformation consists in applying on each element of the matrix $S$ an elementary transformation $s$. We denote by $S_{i,Su}$ the state after the $i^{th}$ SubByte.

$$M_4(GF(2^8)) \longrightarrow M_4(GF(2^8))$$

$$S = \begin{pmatrix} S[1] & S[5] & S[9] & S[13] \\ S[2] & S[6] & S[10] & S[14] \\ S[3] & S[7] & S[11] & S[15] \\ S[4] & S[8] & S[12] & S[16] \end{pmatrix} \longmapsto S_{i,Su} = \begin{pmatrix} s(S[1]) & s(S[5]) & s(S[9]) & s(S[13]) \\ s(S[2]) & s(S[6]) & s(S[10]) & s(S[14]) \\ s(S[3]) & s(S[7]) & s(S[11]) & s(S[15]) \\ s(S[4]) & s(S[8]) & s(S[12]) & s(S[16]) \end{pmatrix},$$

where $s$ is the non linear application defined by

$$\begin{array}{rcl} GF(2^8) & \longrightarrow & GF(2^8) \\ x & \longmapsto & s(x) = \begin{cases} a * x^{-1} + b, & \text{if } x \neq 0, \\ b, & \text{if } x = 0. \end{cases} \end{array}$$

$a$ is a linear invertible application over $GF(2)$, $a \in M_8(GF(2))$, $*$ is the multiplication of matrices over $GF(2)$ and $x^{-1} = \{b_0 b_1 ... b_7\}_b$ is seen as a $GF(2)$-vector equal to tranpose of the vector $(b_0, \cdots, b_7)$. The value of $b = $ '63'$\in GF(2^8)$ and

$$a = \begin{pmatrix} 1 & 0 & 0 & 0 & 1 & 1 & 1 & 1 \\ 1 & 1 & 0 & 0 & 0 & 1 & 1 & 1 \\ 1 & 1 & 1 & 0 & 0 & 0 & 1 & 1 \\ 1 & 1 & 1 & 1 & 0 & 0 & 0 & 1 \\ 1 & 1 & 1 & 1 & 1 & 0 & 0 & 0 \\ 0 & 1 & 1 & 1 & 1 & 1 & 0 & 0 \\ 0 & 0 & 1 & 1 & 1 & 1 & 1 & 0 \\ 0 & 0 & 0 & 1 & 1 & 1 & 1 & 1 \end{pmatrix}.$$

## 2.5 MixColumn for $i^{th}$ round

The MixColumn transformation consists of a multiplication of matrices in $M_4(GF(2^8))$, between the state and a fixed matrix $A_0$ of $M_4(GF(2^8))$. We denote by $S_{i,M}$ the state after the $i^{th}$ MixColumn.

$$\begin{array}{rcl} M_4(GF(2^8)) & \longrightarrow & M_4(GF(2^8)) \\ S & \longmapsto & S_{i,M} = A_0.S, \end{array}$$

where $A_0$ is defined by

$$A_0 = \begin{pmatrix} 02 & 03 & 01 & 01 \\ 01 & 02 & 03 & 01 \\ 01 & 01 & 02 & 03 \\ 03 & 01 & 01 & 02 \end{pmatrix}.$$

## 2.6 ShiftRows for $i^{th}$ round

The ShiftRows transformation is a byte transposition that cyclically shifts the rows of the state over different offsets. We denote by $S_{i,Sh}$ the state after the $i^{th}$ ShiftRows.

$$S = \begin{pmatrix} S[1] & S[5] & S[9] & S[13] \\ S[2] & S[6] & S[10] & S[14] \\ S[3] & S[7] & S[11] & S[15] \\ S[4] & S[8] & S[12] & S[16] \end{pmatrix} \begin{array}{c} M_4(GF(2^8)) \longrightarrow M_4(GF(2^8)) \\ \longmapsto \end{array} S_{i,Sh} = \begin{pmatrix} S[1] & S[5] & S[9] & S[13] \\ S[6] & S[10] & S[14] & S[2] \\ S[11] & S[15] & S[3] & S[7] \\ S[16] & S[4] & S[8] & S[12] \end{pmatrix}.$$

# 3 The description of the attack on computation of AES

First, we are going describe an attack on AES in a simple case and after that we will see how we can generalize this attack. The goal of the attack is to recover the key $K_{Nr}$. Once we discover the subkey $K_{Nr}$, it is easy to get the key $K$, see appendix A.

## 3.1 Principle of the attack

We suppose that we can change a single byte of the state after the ShiftRow of the $N_r - 1$ round and we know the index of the faulty element of state (this last supposition can be omitted, it is more easier to explain the mechanism). The new value of the element of the state is supposed unknown. The fault $\varepsilon$ is spread over four bytes on the output state. For each modified elements on the output state, we find a set of possible fault $\varepsilon$. Moreover we can intersect the possible values $\varepsilon$ for these four sets, we obtain a small set thus reducing the number of required ciphertext for the full analysis. Finally for each fault, we deduce some possible values of four elements of the last roundkey. Repeating ciphertexts, we find four bytes of roundkey 10.

This attack still works out, even with more general assumptions on the fault locations, such as faults without knowing the fault locations before the $9^{th}$ MixColumn transformation. We also expect that faults in round 8 (before the $8^{th}$ MixColumn transformation) might be useful for the analysis, thus growing the number of required ciphertext for the full analysis. With our example, we need ten ciphertexts to get four bytes of roundkey 10, when we don't make hypothesis about the fault locations.

## 3.2 Example

We use the same example as Appendix B of [1]. The following diagram shows the values in the final States array as the Cipher progresses for a block length and a Cipher Key length of 16 bytes each (i.e., Nb = 4 and Nk = 4).

`Input=    '32 43 F6 A8 88 5A 30 8D 31 31 98 A2 E0 37 07 34'`

`Cipher Key=   '2B 7E 15 16 28 AE D2 A6 AB F7 15 88 09 CF 4F 3C'`

`Output=   '39 25 84 1D 02 DC 09 FB DC 11 85 97 19 6A 0B 32'`

The fault propagation appears in grey tint and in hexadecimal notation:

After ShiftRows 9

| 87 | F2 | 4D | 97 |
|----|----|----|----|
| 6E | 4C | 90 | EC |
| 46 | E7 | 4A | C3 |
| A6 | 8C | D8 | 95 |

Fault injected 1E

| 99 | F2 | 4D | 97 |
|----|----|----|----|
| 6E | 4C | 90 | EC |
| 46 | E7 | 4A | C3 |
| A6 | 8C | D8 | 95 |

After Mixcolumn

| 7B | 40 | A3 | 4C |
|----|----|----|----|
| 29 | D4 | 70 | 9F |
| 8A | E4 | 3A | 42 |
| CF | A5 | A6 | BC |

$\oplus$

$K_9$

| AC | 19 | 28 | 57 |
|----|----|----|----|
| 77 | FA | D1 | 5C |
| 66 | DC | 29 | 00 |
| F3 | 21 | 41 | 6E |

After AddRoundKey 9

| D7 | 59 | 8B | 1B |
|----|----|----|----|
| 5E | 2E | A1 | C3 |
| EC | 38 | 13 | 42 |
| 3C | 84 | E7 | D2 |

After SubBytes 10

| 0E | CB | 3D | AF |
|----|----|----|----|
| 58 | 31 | 32 | 2E |
| CE | 07 | 7D | 2C |
| EB | 5F | 94 | B5 |

After ShiftRows 10

| 0E | CB | 3D | AF |
|----|----|----|----|
| 31 | 32 | 2E | 58 |
| 7D | 2C | CE | 07 |
| B5 | EB | 5F | 94 |

$\oplus$

value of $K_{10}$

| D0 | C9 | E1 | B6 |
|----|----|----|----|
| 14 | EE | 3F | 63 |
| F9 | 25 | 0C | 0C |
| A8 | 89 | C8 | A6 |

| Output with Faults |    |    |    |
|---|---|---|---|
| DE | 02 | DC | 19 |
| 25 | DC | 11 | 3B |
| 84 | 09 | C2 | 0B |
| 1D | 62 | 97 | 32 |

The injected error in the state, give four errors in the final state.

## 3.3 How the injected error acts on the final state

We denote by $F$ the faulty state. Now we describe each step from the $N_r - 1^{th}$ MixColumn to the end, and assume that we replace the first element of the state by an unknown value. Let $\varepsilon \in GF(2^8) - \{0\}$ defined by

$$F_{N_r-1,Sh}[1] = S_{N_r-1,Sh}[1] + \varepsilon.$$

### 3.3.1 Fault modification

Obviously

$$F_{N_r-1,Sh} = S_{N_r-1,Sh} + \begin{pmatrix} \varepsilon & 0 & 0 & 0 \\ 0 & 0 & 0 & 0 \\ 0 & 0 & 0 & 0 \\ 0 & 0 & 0 & 0 \end{pmatrix}.$$

### 3.3.2 Effect on MixColumn

$$F_{N_r-1,M} = S_{N_r-1,M} + A_0 \cdot \begin{pmatrix} \varepsilon & 0 & 0 & 0 \\ 0 & 0 & 0 & 0 \\ 0 & 0 & 0 & 0 \\ 0 & 0 & 0 & 0 \end{pmatrix} = S_{N_r-1,M} + \begin{pmatrix} 2.\varepsilon & 0 & 0 & 0 \\ \varepsilon & 0 & 0 & 0 \\ \varepsilon & 0 & 0 & 0 \\ 3.\varepsilon & 0 & 0 & 0 \end{pmatrix}.$$

### 3.3.3 Effect on AddRoundKey

$$F_{N_r-1,A} = S_{N_r-1,A} + \begin{pmatrix} 2.\varepsilon & 0 & 0 & 0 \\ \varepsilon & 0 & 0 & 0 \\ \varepsilon & 0 & 0 & 0 \\ 3.\varepsilon & 0 & 0 & 0 \end{pmatrix}.$$

### 3.3.4 Effect on last SubBytes

We can define $\varepsilon'_0$, $\varepsilon'_1$, $\varepsilon'_2$, $\varepsilon'_3$ (the differential faults) by the following equation

$$F_{N_r,Su} = S_{N_r,Su} + \begin{pmatrix} \varepsilon'_0 & 0 & 0 & 0 \\ \varepsilon'_1 & 0 & 0 & 0 \\ \varepsilon'_2 & 0 & 0 & 0 \\ \varepsilon'_3 & 0 & 0 & 0 \end{pmatrix}.$$

### 3.3.5 Effect after last ShiftRows

$$F_{N_r,Sh} = S_{N_r,Sh} + \begin{pmatrix} \varepsilon'_0 & 0 & 0 & 0 \\ 0 & 0 & 0 & \varepsilon'_1 \\ 0 & 0 & \varepsilon'_2 & 0 \\ 0 & \varepsilon'_3 & 0 & 0 \end{pmatrix}.$$

### 3.3.6 Effect after last AddRoundKey

$$F_{N_r,A} = S_{N_r,A} + \begin{pmatrix} \varepsilon'_0 & 0 & 0 & 0 \\ 0 & 0 & 0 & \varepsilon'_1 \\ 0 & 0 & \varepsilon'_2 & 0 \\ 0 & \varepsilon'_3 & 0 & 0 \end{pmatrix}.$$

$F_{N_r,A}$ is the faulty output for a cipher. Comparing the states $F_{N_r,A}$ with $S_{N_r,A}$, it is easy to get the values of $\varepsilon'_0$, $\varepsilon'_1$, $\varepsilon'_2$ and $\varepsilon'_3$.

## 3.4 Example

Always, in hexadecimal notation, we find

<table>
<tr><td colspan="4">Output with faults</td><td></td><td colspan="4">Output without fault</td><td></td><td colspan="4">Error</td></tr>
<tr><td>DE</td><td>02</td><td>DC</td><td>19</td><td rowspan="4">⊕</td><td>39</td><td>02</td><td>DC</td><td>19</td><td rowspan="4">=</td><td>E7</td><td>00</td><td>00</td><td>00</td></tr>
<tr><td>25</td><td>DC</td><td>11</td><td>3B</td><td>25</td><td>DC</td><td>11</td><td>6A</td><td>00</td><td>00</td><td>00</td><td>51</td></tr>
<tr><td>84</td><td>09</td><td>C2</td><td>0B</td><td>84</td><td>09</td><td>85</td><td>0B</td><td>00</td><td>00</td><td>47</td><td>00</td></tr>
<tr><td>1D</td><td>62</td><td>97</td><td>32</td><td>1D</td><td>FB</td><td>97</td><td>32</td><td>00</td><td>99</td><td>00</td><td>00</td></tr>
</table>

The differential faults are $\varepsilon'_0 =$ 'E7', $\varepsilon'_1 =$ '51', $\varepsilon'_2 =$ '47' and $\varepsilon'_3 =$ '99'.

## 3.5 Analysis on information brought by fault

The only operation that could bring information about the key $K_{N_r}$ is the last SubBytes transformation. Consequently we have four equations where $x_0, x_1, x_2, x_3, \varepsilon$ are unknown variables. We want to solve the following equations (in $x_i$ and $\varepsilon$) :

$$\begin{cases} s(x_0 + 2.\varepsilon) &= s(x_0) + \varepsilon'_0 \\ s(x_1 + \varepsilon) &= s(x_1) + \varepsilon'_1 \\ s(x_2 + \varepsilon) &= s(x_2) + \varepsilon'_2 \\ s(x_3 + 3.\varepsilon) &= s(x_3) + \varepsilon'_3 \end{cases}$$

All these equations belong to a generalized equation

$$s(x + c.\varepsilon) + s(x) = \varepsilon', \qquad (1)$$

where $c =$ '01', '02' or '03' and let us analyse it.

**Definition 1** *We define the set of solutions of (1) in $\varepsilon$ by*

$$S_{c,\varepsilon'} = \left\{ \varepsilon \in GF(2^8) : \exists x \in GF(2^8), \ s(x + c.\varepsilon) + s(x) = \varepsilon' \right\}.$$

**Definition 2** *Consider the linear application over $GF(2)$:*

$$\begin{array}{rcl} l: GF(2^8) &\longrightarrow& GF(2^8) \\ x &\longmapsto& x^2 + x \end{array}$$

Denote by $E_1 = Im(l)$ be the $GF(2)$-vector space image of $l$ and $\dim_{GF(2)}(E_1) = 7$. If $\theta \in E_1$, then there are two solutions $x_1, x_2 \in GF(2^8)$ of equation $x^2 + x = \theta$, and the solutions verify $x_2 = x_1 + 1$.

**Definition 3** *Let $\lambda \in GF(2^8)$, $\lambda \neq 0$ and define $\phi_\lambda$ an isomorphism of $GF(2)$-vector spaces*

$$\begin{array}{rcl} \phi_\lambda: GF(2^8) &\longrightarrow& GF(2^8) \\ x &\longmapsto& \lambda.x \end{array}$$

and let $E_\lambda = Im(\phi_\lambda|_{E_1})$ be the $GF(2)$-vector space image of $\phi_\lambda$ restricted to $E_1$. Moreover $\dim_{GF(2)}(E_\lambda) = 7$.

**Proposition 1** *There is a bijective application $\phi$ between $E_1^*(= E_1 - \{0\})$ and $S_{c,\varepsilon'}$.*

$$\begin{array}{rcl} \phi: E_1^* &\longrightarrow& S_{c,\varepsilon'} \\ t &\longmapsto& (c(a^{-1} * \varepsilon').t)^{-1}. \end{array}$$

$S_{c,\varepsilon'}$ have 127 elements.

<u>Proof</u>: Let $\varepsilon \in S_{c,\varepsilon'}$, then $\exists x \in GF(2^8)$ such that (1) holds.
Assume $x \neq 0$ and $x \neq c.\varepsilon$, we get

$$x^2 + c.\varepsilon.x = (a^{-1} * \varepsilon')^{-1}.c.\varepsilon.$$

We denote by $t = x.(c.\varepsilon)^{-1} \in GF(2^8) - \{0\}$, then we have

$$t^2 + t = (a^{-1} * \varepsilon')^{-1}.(c.\varepsilon)^{-1}. \qquad (2)$$

Therefore $(a^{-1} * \varepsilon')^{-1}(c.\varepsilon)^{-1} \in E_1^*$. Reciprocally for $\theta \in E_1^*$ we can define $(a^{-1} * \varepsilon')^{-1}.(c.\theta)^{-1} \in S_{c,\varepsilon'}$.
Assume $x = 0$ or $x = c.\varepsilon$, (1) becomes $a * (c.\varepsilon)^{-1} = \varepsilon'$. We obtain $\varepsilon = ((a^{-1} * \varepsilon').c)^{-1}$, this case is included in the previous case because $1 \in E_1^*$. We see for the case $\theta = 1$, the equation (1) has four solutions in $x$. In brief, there exists a bijection map between $E_1^*$ and $S_{c,\varepsilon'}$:

$$E_1^* \xrightarrow{\phi_\lambda} E_\lambda - \{0\} \longrightarrow S_{c,\varepsilon'}$$
$$t \longmapsto \lambda.t \longmapsto (\lambda.t)^{-1}.$$

where $\lambda = c(a^{-1} * \varepsilon')$.

□

**Proposition 2** *The following statements hold for $\lambda_1, \lambda_2 \in GF(2^8) - \{0\}$:*

$$dim_{GF(2)}(E_{\lambda_1} \cap E_{\lambda_2}) = \begin{cases} 7 & \text{If } \lambda_1 = \lambda_2 \\ 6 & \text{Otherwise} \end{cases}$$

Proof: This proof comes from the following lemma :

□

**Lemma 1** *For $\lambda_1, \lambda_2 \in GF(2^8) - \{0\}$, we get*

$$E_{\lambda_1} = E_{\lambda_2} \iff \lambda_1 = \lambda_2.$$

Proof: This lemma is equivalent to this assertion : for $\lambda \in GF(2^8) - \{0\}$,

$$E_\lambda = E_1 \iff \lambda = 1.$$

Let us prove this statement and assume that $\lambda E_1 = E_1$. Remark that $E_1 = \{e = \{e_7 e_6 \cdots e_0\}_b \in GF(2^8) - \{0\} : e_7 = e_5\}$. Hence $\{1, x, x^2, x^3, x^4, x^6, x^5 + x^7\}$ is a basis of $E_1$. Multiply the basis's vectors $v_i$ with $\lambda = \{\lambda_7 \cdots \lambda_0\}_b$. As $\lambda v_i \in E_1$, we have $(\lambda v_i)_7 = (\lambda v_i)_5$. We obtain 7 relations ($\lambda_7 = \lambda_5$, $\lambda_6 = \lambda_4$, $\lambda_5 = \lambda_3 + \lambda_7$, $\lambda_4 = \lambda_6 + \lambda_2 + \lambda_7$, $\lambda_7 + \lambda_3 = \lambda_5 + \lambda_1 + \lambda_6$, $\lambda_5 + \lambda_1 = \lambda_3 + \lambda_4$, $\lambda_6 + \lambda_5 = \lambda_7 + \lambda_3$). We solve this system to obtain $\lambda_7 = \lambda_6 = \lambda_5 = \lambda_4 = \lambda_3 = \lambda_2 = \lambda_1 = 0$. The solution $\lambda = 0$ doesn't match. We have $\lambda = 1$. □

**Proposition 3** *For $\lambda_1, \lambda_2, \lambda_3 \in GF(2^8) - \{0\}$, we get:*

$$dim_{GF(2)}(E_{\lambda_1} \cap E_{\lambda_2} \cap E_{\lambda_3}) = \begin{cases} 7 & \text{If } \lambda_1 = \lambda_2 = \lambda_3 \\ 6 & \text{If } rank_{GF(2)}\{\lambda_1^{-1}, \lambda_2^{-1}, \lambda_3^{-1}\} = 2 \\ 5 & \text{Otherwise} \end{cases}$$

Proof: It comes from proposition 2 and this following lemma

□

**Lemma 2** *For $\lambda_1, \lambda_2, \lambda_3 \in GF(2^8) - \{0\}$, we get*

$$E_{\lambda_1} \cap E_{\lambda_3} = E_{\lambda_2} \cap E_{\lambda_3} \iff \lambda_3^{-1} = \lambda_1^{-1} + \lambda_2^{-1} \text{ or } \lambda_1 = \lambda_2.$$

Proof:

1. $\Leftarrow$
   Let $x \in E_{\lambda_1} \cap E_{\lambda_3}$, then $\exists y, t \in E_1$ such that $x = \lambda_1.y = \lambda_3.t$.

   $$y = \lambda_1^{-1}.\lambda_3.t = \lambda_2^{-1}.\lambda_3.t + t,$$

   $$y - t = \lambda_2^{-1}.\lambda_3.t \in E_1,$$

   and

   $$x = \lambda_3.t = \lambda_2.(y - t) \in E_{\lambda_2}$$

2. $\Rightarrow$
   Assume that $\lambda_1 \neq \lambda_2$, and let us show $\forall t \in E_1$, $\lambda_3.(\lambda_1^{-1} + \lambda_2^{-1}).t \in E_1$.
   Let $x = \lambda_3.t \in E_{\lambda_3}$:

   - If $x \in E_{\lambda_1}$ then $x \in E_{\lambda_2}$ therefore $\exists s_1, s_2 \in E_1$ such that $x = \lambda_1.s_1 = \lambda_2.s_2$ and we get $\lambda_3.(\lambda_1^{-1} + \lambda_2^{-1}).t = s_1 + s_2 \in E_1$.
   
   - If $x \notin E_{\lambda_1}$ then $x \notin E_{\lambda_2}$ therefore we get $\lambda_1^{-1}.x \notin E_1$ and $\lambda_2^{-1}.x \notin E_1$. We have $\lambda_3.(\lambda_1^{-1} + \lambda_2^{-1}).t = \lambda_1^{-1}.x + \lambda_2^{-1}.x \in E_1$ (because $\forall u \notin E_1$ and $\forall v \notin E_1$ then $u + v \in E_1$).

We showed that $E_{\lambda_3.(\lambda_1^{-1}+\lambda_2^{-1})} = E_1$ and with the lemma 1 we get $\lambda_3^{-1} = \lambda_1^{-1} + \lambda_2^{-1}$.

□

**Proposition 4** *Finally for $\lambda_1, \lambda_2, \lambda_3, \lambda_4 \in GF(2^8) - \{0\}$, we get:*

$$dim_{GF(2)}(E_{\lambda_1} \cap E_{\lambda_2} \cap E_{\lambda_3} \cap E_{\lambda_4}) = \begin{cases} 7 & \text{If } \lambda_1 = \lambda_2 = \lambda_3 = \lambda_4 \\ 6 & \text{If } rank_{GF(2)}\{\lambda_1^{-1}, \lambda_2^{-1}, \lambda_3^{-1}, \lambda_4^{-1}\} = 2 \\ 5 & \text{If } rank_{GF(2)}\{\lambda_1^{-1}, \lambda_2^{-1}, \lambda_3^{-1}, \lambda_4^{-1}\} = 3 \\ 4 & \text{Otherwise} \end{cases}$$

**Definition 4** *We considered four equations in a different way, but the committed fault is common to these four equations, that is why we introduce the set of possible committed faults $S$ :*

$$S = S_{2,\varepsilon'_0} \bigcap S_{1,\varepsilon'_1} \bigcap S_{1,\varepsilon'_2} \bigcap S_{3,\varepsilon'_3}.$$

*Moreover the cardinal of $S$ is smaller than the cardinal of $S_{c,\varepsilon}$. It allows to reduce the space of the faults, and so to use fewer faulty calculations to go back up to the key.*

**Corollary 1** *If two of the four following values $2^{-1}.\varepsilon'_0$, $\varepsilon'_1$, $\varepsilon'_2$, $3^{-1}.\varepsilon'_3$ are not equal, we have*

$$Card\ (S_{2,\varepsilon'_0} \bigcap S_{1,\varepsilon'_1} \bigcap S_{1,\varepsilon'_2} \bigcap S_{3,\varepsilon'_3}) \leq 63.$$

**Proposition 5** *For a differential fault $\varepsilon'$, let $\varepsilon \in S \cap S_{c,\varepsilon'}$ be a fault value and define $\theta = ((a^{-1} * \varepsilon').c.\varepsilon)^{-1} \in E_1^*$ and $\alpha, \beta$ the two solutions (in $GF(2^8)$) of the equation $t^2 + t = \theta$. The possible values of key $K_{N_r}[i]$ (for some $i$, it is the index of element in the state) are*

- *If $\theta \neq 1$, then there are two possible values of $K_{N_r}[i]$*

$$K_{N_r}[i] = s(c.\varepsilon.\alpha) + F_{N_r,A}[i] \text{ or } K_{N_r}[i] = s(c.\varepsilon.\beta) + F_{N_r,A}[i]$$

- *If $\theta = 1$, then there are four possible values of $K_{N_r}[i]$*

$$K_{N_r}[i] = s(c.\varepsilon.\alpha) + F_{N_r,A}[i] \text{ or } K_{N_r}[i] = s(c.\varepsilon.\beta) + F_{N_r,A}[i]$$

$$\text{or } K_{N_r}[i] = b + F_{N_r,A}[i] \text{ or } K_{N_r}[i] = s(c.\varepsilon) + F_{N_r,A}[i]$$

Proof:

- If $\theta \neq 1$, we know that $\theta \in E_1$, then there are two solutions $\alpha, \beta$ of $t^2 + t = \theta$. We deduce two solutions from (1) noted $\{x_1, x_2\}$, by $x_1 = c.\varepsilon.\alpha$ and $x_2 = c.\varepsilon.\beta$.

- If $\theta = 1$, we know that $1 \in E_1$, then there are two solutions $\alpha, \beta$ of $t^2 + t = 1$. We deduce two solutions from (1) noted $\{x_1, x_2\}$, by $x_1 = c.\varepsilon.\alpha$ and $x_2 = c.\varepsilon.\beta$. Moreover there are also two trivial solutions of (1) : $x_3 = 0$ and $x_4 = c.\varepsilon$.

Once we get a solution $x$ of (1), it is easy to get a possible value of $K_{N_r}[i]$. □

By applying this proposition to the four faulty elements of the state, we can deduce four sets of possible values for $K_{N_r}[0]$, $K_{N_r}[7]$, $K_{N_r}[10]$ and $K_{N_r}[13]$. Then by repeating the insertion of faults in a calculation, and by intersecting these four sets we get rather quickly a single value for $K_{N_r}[0]$, $K_{N_r}[7]$, $K_{N_r}[10]$ and $K_{N_r}[13]$.

### 3.6 Example

Remember our example:

$$\begin{aligned} s(x_0 \oplus 2.\varepsilon) &= s(x_0) \oplus \text{'E7'} \\ s(x_1 \oplus \varepsilon) &= s(x_1) \oplus \text{'51'} \\ s(x_2 \oplus \varepsilon) &= s(x_2) \oplus \text{'47'} \\ s(x_3 \oplus 3.\varepsilon) &= s(x_3) \oplus \text{'99'} \end{aligned}$$

Let $E_1 = \{\text{'01'..'1F','40'..'5F','A0'..'BF','E0'..'FF'}\}$ and $S_{c,\varepsilon'} = \{(c.(a^{-1} * \varepsilon').e)^{-1},\ e \in E1\}$.

We compute

$$S_{2,'E7'} \bigcap S_{1,'51'} \bigcap S_{1,'47'} \bigcap S_{3,'99'}$$
$$= \{'01', '04', '13', \textbf{'1E'}, '21', '27', '33', '3B', '48', '4D', '50', '53', '55', '5D', '64', '65',$$
$$'7E', '7F', '80', '83', '8D', '8F', '93', 'A7', 'A8', 'A9', 'AB', 'B3', 'B8', 'C9', 'F6'\}$$

We get (the real value of $K_{10}[0]$ is 'D0')

$K_{10}[0] \in \{$'03', '06', '09', '0C', '10', '15', '1A', '1F', '21', '24', '2B', '2E', '32', '37', '38', '3D', '43', '46', '49', '4C', '50', '55', '5F', '61', '64', '6B', '6E', '72', '77', '78', '7D', '83', '86', '89', '8C', '90', '95', '9A', '9F', 'A1', 'A4', 'AB', 'AE', 'B2', 'B7', 'B8', 'C3', 'C6', 'C9', 'CC', **'D0'**, 'D5', 'DA', 'DF', 'E1', 'E4', 'EB', 'EE', 'F2', 'F7', 'F8', 'FD'$\}$

With the five faults {'1E', 'E1', 'B3', '16', '9E'}, we obtain a correct and single value of $K_{10}[0]$, $K_{10}[7]$, $K_{10}[10]$, $K_{10}[13]$.

## 4 Generalisation

### 4.1 Without fault location

In this section, we assume that the fault is on a byte, between the last two MixColumn. It's the same case than previously except that the fault can be confined on the byte 1 to 16. The fault is propagated by the MixColumn and spread on 4 bytes of the state. On the first line of the differential state matrix, we have a induced fault. We can determine from which column the injected fault belongs by considering the column of induced fault. Next we analyse the four possibilities of line position for the injected fault with the method presented in previous section.

### 4.2 Hardware Device

Suppose that you can physically modify an hardware AES device. First, compute ciphers for more than ten random plaintexts with AES device. Next, modify by example the design by cutting lines and connecting them to the earth (or Vcc) temporaly between two bytes during the round located two rounds before the end. It amounts to having a byte of round $N_k - 2$, always replaced by '00' (or 'FF'). Compute an other time the same messages with the tampered device. With random plaintexts, the faulty byte is like an random fault. This fault is passed on four faults at round $N_k - 1$ and sixteen faults at round $N_k$. It is this differential matrix we can analyse error by error to find the last round key.

## A Back to initial key with the last subkey

See [1] for additional informations about $w$ and RotWord, Rcon and SubWord functions.

Let us denote by $K_n[j]$ the $j^{th}$ byte of the $n^{th}$ roundkey and $w[i]$ as in [1]. We have

$$K_n = (w[N_k n], w[N_k n + 1], \cdots, w[N_k n + N_k - 1]).$$

We have the following relations (for $N_k = 4, 6$):
for $N_k \leqslant i < Nb * (Nr + 1), i \neq 0 \mod N_k$,

$$w[i] = w[i - N_k] \oplus w[i - 1]$$
$$\text{i.e. } w[i - N_k] = w[i] \oplus w[i - 1]$$

and for $i = 0 \mod N_k$,

$$w[i] = w[i - N_k] \oplus \text{SubWord}(\text{RotWord}(w[i - 1])) \oplus \text{Rcon}[i/N_k]$$
$$\text{i.e. } w[i - N_k] = w[i] \oplus \text{SubWord}(\text{RotWord}(w[i - 1])) \oplus \text{Rcon}[i/N_k]$$

Hence, we have
for $0 \leqslant i < Nb * (Nr + 1) - Nk, i \neq 0 \mod N_k$,

$$w[i] = w[i + N_k] \oplus w[i + N_k - 1] \tag{3}$$

and for $i = 0 \mod N_k$,

$$w[i] = w[i+N_k] \oplus \text{SubWord}(\text{RotWord}(w[i+N_k-1])) \oplus \text{Rcon}[(i+N_k)/N_k] \quad (4)$$

With AES-256, you must add an Subword operation when $i \equiv 4 \mod N_k$. So we can deduce previous key from the ending subkey and step by step obtain $K_0$ with is the cipherkey.

```
RecoverKey(byte Finalkey[4*Nk], word w[Nb*(Nr+1)], Nk)
begin
    word temp
    i = Nb * (Nr+1)-1
    j = Nk - 1
    while (j >= 0)
       w[i] = word(Finalkey[4*j], Finalkey[4*j+1],
                   Finalkey[4*j+2], Finalkey[4*j+3])
       i = i-1
       j = j-1
    end while
 {here, "i" must be equal to Nb * (Nr+1) - Nk - 1}
    while (i >= 0)
        temp = w[i+Nk-1]
        if (i mod Nk = 0)
            temp = SubWord(RotWord(temp)) xor Rcon[i/Nk+1]
        else if (Nk > 6 and i mod Nk = 4)
            temp = SubWord(temp)
        end if
        w[i] = w[i+Nk] xor temp
        i = i - 1
    end while
end
```

Figure 1: Pseudo Code for Key Recovery.

**Remark 1** *On AES-128, it is sufficient to know $K_{10}$ to find the cipher key, but on AES-256, you must know $K_{13}$ and $K_{14}$.*